\documentstyle[aps, epsf, psfig]{revtex}

\newcommand{\ep}{\epsilon}

\newcommand{\pa}{\partial}

\newcommand{\td}{\tilde}
\newcommand{\sla}[1]{\slash\!\!\! #1}
\begin{document}

\draft
\title{Rigorous Effective Field Theory Study on Pion Form Factor}
\author{Xiao-Jun Wang\footnote{E-mail address: wangxj@mail.ustc.edu.cn}}
\address{Center for Fundamental Physics,
University of Science and Technology of China\\
Hefei, Anhui 230026, P.R. China}
\author{Mu-Lin Yan{E-mail address: mlyan@staff.ustc.edu.cn}}
\address{CCST(World Lad), P.O. Box 8730, Beijing, 100080, P.R. China \\
  and \\
Center for Fundamental Physics,
University of Science and Technology of China\\
Hefei, Anhui 230026, P.R. China\footnote{mail
address}}
\date{\today}
\maketitle

\begin{abstract}
{We study $e^+e^-\rightarrow\pi^+\pi^-$ cross section and phase shift of
$I=l=1$ $\pi-\pi$ scattering below 1GeV in framework of chiral constituent
quark model. The results including all order contribution of the chiral
perturbation expansion and all one-loop effects of pseudoscalar
mesons, but without any adjust parameters. Width of $\rho$ predicted by
the model strongly depends on transition momentum-square $q^2$.
We show that the mass pamameter of $\rho$-meson in its propagator is very
different from its physical mass due to momentum-dependent width of
$\rho$. The mass difference between $\rho^0$ and $\omega$ are predicted
successfully. The rigorous theoretical prediction on
$e^+e^-\rightarrow\pi^+\pi^-$ cross section and the phase shift in $I=l=1$
$\pi-\pi$ scattering agree with data excellentlly.}
\end{abstract}

\section*{}
The process $e^+e^-\rightarrow\pi^+\pi^-$ at energies lower than the 
chiral symmetry spontaneously breaking scale contains very important
information on low energy hadron dynamics. It was an active subjuct
and studied continually during past fifty years. Experimentally,
the effects of the strong interaction in process of $e^+e^-$
annihilation is obvious to provides a large enhancement to production of
pions in vector meson resonance region[1-5]. Theorectically, however, the
problem was not studied by using a rigorous effective field theory(EFT) of
QCD yet. Although at very low energy the chiral perturbative
theory(ChPT) is a rigorou EFT of QCD, it in principle can not
predict physics at vector meson resonance region. From viewpoint of
quantum field theory, a rigorous theoretical study on
$e^+e^-\rightarrow\pi^+\pi^-$ cross section and $l=1,\;I=1$ $\pi-\pi$
phase shift provided by an EFT of QCD must satisfy the following
requirements: 1) Some fundamental principles, such as symmetry and
unitarity, must be satisfied in this EFT. 2) The experimental data of
$l=1,\;I=1$ $\pi-\pi$ scattering
phase shift implies that width of $\rho$-meson $\Gamma_\rho$ is
transitional momentum-dependent, and must vanish at $q^2=0$(where
$q^2$ denotes four-momentuma squre of off-shell $\rho$). The
momentum-dependence of $\Gamma_\rho(q^2)$ should be predicted by
the EFT itself instead of being fitted by experiment. 3) This EFT
must provide an effective method to evaluated error bar of this
current calculation, i.e., the next order contribution should be
able to be calculated. The purpose of this present paper is to
provide a rigorous EFT study on pion form factor and $I=1$, P-wave
$\pi\pi$ phase shift below 1GeV. In other words, all requirements
mentioned above will be met in the study of this paper.

In some recent refrences\cite{Bena93,Bena97,Connell96}, the
authors have studied $e^+e^-\rightarrow\pi^+\pi^-$ cross section
and $l=1,\;I=1$ phase shift at vector meson resonance region by using some
very simple phenomenological models. These
models are constructed in the intermediate energy region using some
phenomenology considerations, such as vector meson dominant(VMD) and
universal coupling. In principle, each of them can capture some
leading order effects of low energy EFT of QCD and they are classfied
by different symmetry realization for vector meson fields\cite{Brise97}.
However, so far, the low energy effective lagrangians including
vector meson resonances are only up to $O(p^4)$ which are not enough for
the physics at vector meson mass scale, and can not successfully
evaluate very important one-loop effects of pseudoscalar mesons which
corresponds to the next to leading order of $N_c^{-1}$ expansion.
Hence, these phenomenological models are not of rigorous EFT, and
they can not provide any rigorous theoretical predictions on low energy
hardon physics. This bad shortage can be overcome by using the EFT in
ref.\cite{Rho}, in which we constructed a consistent chiral constituent
quark model(ChCQM) with lowest vector meson resonances and element
Goldstone bosons. In this formalism we can capture all order information
of chiral perturbative expansion and one-loop effects of pseudoscalar
mesons.

In chiral limit, ChCQM is parameterized by the following
chiral constituent quark lagrangian
\begin{eqnarray}\label{1}
{\cal L}_{\chi}&=&i\bar{q}(\sla{\pa}+\sla{\Gamma}+
  g_{_A}{\slash\!\!\!\!\Delta}\gamma_5-i\sla{V})q-m\bar{q}q
   +\frac{F^2}{16}<\nabla_\mu U\nabla^\mu U^{\dag}>
   +\frac{1}{4}m_0^2<V_\mu V^{\mu}>.
\end{eqnarray}
Here $<...>$ denotes trace in SU(3) flavour space,
$\bar{q}=(\bar{q}_u,\bar{q}_d,\bar{q}_s)$ are constituent quark
fields. $V_\mu$ denotes vector meson octet and singlet,
\begin{equation}
\label{2}
   V_\mu(x)={\bf \lambda \cdot V}_\mu =\sqrt{2}
\left(\begin{array}{ccc}
       \frac{\rho^0_\mu}{\sqrt{2}}+\frac{\omega_\mu}{\sqrt{2}}
            &\rho^+_\mu &K^{*+}_\mu   \\
    \rho^-_\mu&-\frac{\rho^0_\mu}{\sqrt{2}}+\frac{\omega_\mu}{\sqrt{2}}
            &K^{*0}_\mu   \\
       K^{*-}_\mu&\bar{K}^{*0}_\mu&\phi_\mu
       \end{array} \right).
\end{equation}
The $3\times 3$ anti-Hermian matrices $\Delta_\mu$ and $\Gamma_\mu$
are defined as follows,
\begin{eqnarray}\label{3}
\Delta_\mu&=&\frac{1}{2}\{\xi^{\dag}(\pa_\mu-ir_\mu)\xi
          -\xi(\pa_\mu-il_\mu)\xi^{\dag}\}, \nonumber \\
\Gamma_\mu&=&\frac{1}{2}\{\xi^{\dag}(\pa_\mu-ir_\mu)\xi
          +\xi(\pa_\mu-il_\mu)\xi^{\dag}\},
\end{eqnarray}
and covariant derivative are defined as follows
\begin{eqnarray}\label{4}
\nabla_\mu U&=&\pa_\mu U-ir_\mu U+iUl_\mu=2\xi\Delta_\mu\xi,
  \nonumber \\
\nabla_\mu U^{\dag}&=&\pa_\mu U^{\dag}-il_\mu U^{\dag}+iU^{\dag}r_\mu
  =-2\xi^{\dag}\Delta_\mu\xi^{\dag},
\end{eqnarray}
where $l_\mu=v_\mu+a_\mu$ and $r_\mu=v_\mu-a_\mu$ are linear combinations
of external vector field $v_\mu$ and axial-vector field $a_\mu$, $\xi$
associates with non-linear realization of spontanoeusly broken global
chiral symmetry introduced by Weinberg\cite{Wein68}. This realization is
obtained by specifying the action of global chiral group $G=SU(3)_L\times
SU(3)_R$ on element $\xi(\Phi)$ of the coset space $G/SU(3)_{_V}$:
\begin{equation}\label{5}
\xi(\Phi)\rightarrow
g_R\xi(\Phi)h^{\dag}(\Phi)=h(\Phi)\xi(\Phi)g_L^{\dag},\hspace{0.5in}
 g_L, g_R\in G,\;\;h(\Phi)\in H=SU(3)_{_V}.
\end{equation}
Explicit form of $\xi(\Phi)$ is usual taken
\begin{equation}\label{6}
\xi(\Phi)=\exp{\{i\lambda^a \Phi^a(x)/2\}},\hspace{1in}
U(\Phi)=\xi^2(\Phi),
\end{equation}
where the Goldstone boson $\Phi^a$ are treated as pseudoscalar
meson octet. In ref.\cite{Rho} we have shown that the
lagrangian(~\ref{1}) is invariant under $G_{\rm global}\times
G_{\rm local}$. The axial coupling constant $g_A=0.75$ is fitted by
$\beta$-decay of neutron, and constituent quark mass $m=480$MeV is
fitted by low energy limit of the model. It has been also
illustrated that the value of $g_A$ has included effects of
intermediate axial-vector meson resonances exchanges at low energy.

The EFT describing low energy meson interaction can be deduced via loop
effects of constituent quarks\cite{Rho}. From viewpoint of
symmetry, at leading order of vector mesons coupling to pseudoscalar
mesons, the effective lagrangian is equivalent to WCCWZ lagrangian given
by Brise\cite{Brise97,WCCWZ}. In terms of this EFT, we found that the
chiral perturbative expansion converge slowly at vector
meson energy scale. Thus the high order contributions of chiral
perturbative expansion play important role at this energy scale.
Phenomenologically, this model provides excellent theoretical predictions
on $\rho$-physics\cite{Rho} and on $\omega$-physics\cite{Omega}. In
this present paper, we focus our attention on vector-photon,
vector-$\pi\pi$ and photon-$\pi\pi$ vertices. These relevant
vertices have been calculated in ref.\cite{Rho,Omega} which
including all order effects of the chiral perturbative expansion
and one-loop contribution of pseudoscalar mesons. The ``direct''
photon-$\pi\pi$ coupling and vector-photon coupling vertices read,
\begin{eqnarray}\label{7}
{\cal L}_{\gamma\pi\pi}^c&=&\int\frac{d^4q}{(2\pi)^4}e^{iq\cdot x}
\bar{F}_\pi(q^2)A_\mu(q)[\pi^+(x)\pa^\mu\pi^-(x)-\pa^\mu\pi^+(x)\pi^-(x)],
     \nonumber \\
{\cal L}_{\rho\gamma}^c&=&-\frac{1}{2}e\int\frac{d^4q}{(2\pi)^4}
  e^{iq\cdot x}b_{\rho\gamma}(q^2)(q^2\delta_{\mu\nu}-q_\mu q_\nu)
  \rho^{0\mu}(q)A^\nu(x), \\
{\cal L}_{\omega\gamma}^c&=&-\frac{1}{6}e\int\frac{d^4q}{(2\pi)^4}
  e^{iq\cdot x}b_{\rho\gamma}(q^2)(q^2\delta_{\mu\nu}-q_\mu q_\nu)
  \omega^{\mu}(q)A^\nu(x), \nonumber   
\end{eqnarray}
where the super-srcipt ``c'' denotes these ``complete'' effective 
couplings which have contained one-loop effects of pseudoscalar mesons so
that the ``form factors'', $\bar{F}_\pi(q^2)$, $b_{\rho\gamma}(q^2)$ etc.,
are not real function. These ``form factors'' are given as
follows\cite{Rho,Omega},
\begin{eqnarray}\label{8}
\bar{F}_\pi(q^2)&=&1+\frac{q^2b_\gamma(q^2)}{1+\Sigma(q^2)}, \hspace{1in}
b_{\rho\gamma}(q^2)=\frac{A(q^2)}{g(1+\zeta)}-f_\pi^2b(q^2)
  \frac{\Sigma_0(q^2)}{1+2\zeta}[1
  +\frac{q^2b_\gamma(q^2)}{1+\Sigma(q^2)}], \nonumber\\
b_\gamma(q^2)&=&\frac{gb(q^2)}{2(1+3\zeta)}-
  \frac{1}{16\pi^2f_\pi^2}\{\lambda+\int_0^1dx\cdot
  x(1-x)\ln{[(1-\frac{x(1-x)p^2}{m_{_K}^2})
 (\frac{x(1-x)p^2}{m_{_K}^2})^2]}\nonumber \\&&\hspace{1in}
  -\frac{2}{3}i\pi\theta(p^2-4m_\pi^2)\}
  -\frac{C(q^2)\Sigma_0(q^2)}{(1+11\zeta/3)},
   \nonumber \\
b(q^2)&=&\frac{1}{gf_\pi^2}[A(q^2)+g_A^2B(q^2)],\hspace{1in}
C(q^2)=\frac{A(q^2)+2g_{_A}^2B(q^2)}{2f_\pi^2},\nonumber \\
A(q^2)&=&g^2-\frac{N_c}{\pi^2}\int_0^1 dt\cdot t(1-t)\ln{(1
 -\frac{t(1-t)q^2}{m^2})}, \\
B(q^2)&=&-g^2+\frac{N_c}{2\pi^2}\int_0^1dt_1\cdot t_1\int_0^1dt_2
 (1-t_1t_2)[1+\frac{m^2}{m^2-t_1(1-t_1)(1-t_2)q^2} \nonumber \\
 &&+\ln{(1-\frac{t_1(1-t_1)(1-t_2)q^2}{m^2})}], \nonumber \\
\Sigma_0(q^2)&=&\frac{2}{f_\pi^2}[2\Sigma_\pi(q^2)-\Sigma_K(q^2)],
   \hspace{1in}
\Sigma(q^2)=[1+\frac{q^2C(q^2)}{1+11\zeta/3}]\Sigma_0(q^2),
  \nonumber \\
\Sigma_{K}(q^2)&=&\frac{1}{(4\pi)^2}\{\lambda(m_{_K}^2-\frac{q^2}{6})
  +\int_0^1dt[m_{_K}^2-t(1-t)q^2]\ln{(1-\frac{t(1-t)q^2}{m_{_K}^2})}\},
        \nonumber \\
\Sigma_\pi(q^2)&=&\frac{q^2}{(4\pi)^2}\{\frac{\lambda}{6}
  +\int_0^1dt\cdot t(1-t)\ln{\frac{t(1-t)q^2}{m_{_K}^2}}
  -\frac{i}{6}\pi\theta(q^2-4m_\pi^2)\}, \nonumber 
\end{eqnarray}
where $f_\pi=185$MeV is decay constant of pion, $g$ and $\lambda$ are
constants which absorb the logarithmic divergence from constituent quark
loops and the quadratic divergence from meson loops respectively, 
\begin{eqnarray}\label{9}
g^2&=&\frac{8}{3}\frac{N_c}{(4\pi)^{D/2}}(\frac{\mu^2}{m^2})^{\ep/2}
  \Gamma(2-\frac{D}{2}), \nonumber \\
\lambda&=&(\frac{4\pi\mu^2}{m_{_K}^2})^{\ep/2}\Gamma(2-\frac{D}{2}),
 \hspace{1in}\zeta=\frac{2\lambda}{(4\pi)^2}\frac{m_{_K}^2}{f_\pi^2}.
\end{eqnarray}
In ref.\cite{Rho}, $\lambda=2/3$ has been fitted by Zweig rule.

Traditionally, VMD\cite{VMD} assumes that all photon-hadron coupling is
mediated by vector mesons. However, from an empirical or symmetrical point
of view, one has a freedom, that a non-resonant background is alowed.
For instance, in process of $e^+e^-\rightarrow\pi^+\pi^-$, since
$\pi^+\pi^-$ can consist of a vector-isovector system whose quantum
numbers are same to $\rho^0$, experiment or symmetry can not divide
contribution of ``direct'' photon-$\pi\pi$ coupling from one from
photon$\rightarrow\rho^0\rightarrow\pi^+\pi^-$. Therefore, in general,
the traditional VMD is a strong assumption. From eq.~(\ref{7}), we can
see that the ``direct'' photon-$\pi\pi$ coupling indeed exists in this
EFT. The same problem is also questioned in isospin breaking decay
$\omega\rightarrow\pi^+\pi^-$, which is dominated by $\rho^0$ exchange,
but have a contirbution from ``direct'' $\omega\pi^+\pi^-$ coupling yet.

The ``complete'' $\rho\pi\pi$ vertex reads
\begin{equation}\label{10}
{\cal L}_{\rho\pi\pi}^c=\int
  \frac{d^4q}{(2\pi)^4}e^{iq\cdot x}g_{\rho\pi\pi}(q^2)
  (q^2\delta_{\mu\nu}-q_\mu q_\nu)\rho^{0\mu}(q)
  [\pi^+(x)\pa^\mu\pi^-(x)-\pa^\mu\pi^+(x)\pi^-(x)],
\end{equation}
with
\begin{equation}\label{11}
g_{\rho\pi\pi}(q^2)=\frac{b(q^2)}{(1+2\zeta)(1+\Sigma(q^2))}.   
\end{equation}

At leading order of vector meson coupling, the VMD vertex and $\rho\pi\pi$
vertex read respectively
\begin{eqnarray}\label{12}
{\cal L}_{\rho\gamma}^c&=&-\frac{1}{2}eg\int\frac{d^4q}{(2\pi)^4}
  e^{iq\cdot x}(q^2\delta_{\mu\nu}-q_\mu q_\nu)
  \rho^{0\mu}(q)A^\nu(x), \nonumber \\
{\cal L}_{\rho\pi\pi}^c&=&\int\frac{d^4q}{(2\pi)^4}e^{iq\cdot x}
  \frac{1}{gf_\pi^2}[g^2+g_{_A}^2(\frac{N_c}{3\pi^2}-g^2)]
  (q^2\delta_{\mu\nu}-q_\mu q_\nu)\rho^{0\mu}(q)
  [\pi^+(x)\pa^\mu\pi^-(x)-\pa^\mu\pi^+(x)\pi^-(x)].
\end{eqnarray}
A gauge-like argument\cite{VMD,Connell97} suggests that the $\rho$ couples
to all hadrons with the same strength(universality). It is formulated by
the first KSRF sum rule\cite{KSRF}
\begin{eqnarray}\label{13}
g_{\rho\gamma}=\frac{1}{2}f_\pi^2g_{\rho\pi\pi}.
\end{eqnarray}
However, experimentally, the first KSRF sum rule is observed to be not
quite exact\cite{KKW96}. It can be naturally understood the universal
coupling is a conclusion only working at leading order of vector meson
coupling, and high order contribution will correct it. From eq.~(\ref{12})
we can see that the first KSRF sum rule is strictly satified when
$g=\pi^{-1}$ for $N_c=3$. Thus $g=\pi^{-1}$ is a favorite choice. In
addition, it has been shown in ref.\cite{Rho} how high order correction
breaks the first KSRF sum rules. Besides of the parameters $g_A$, $m$, $g$
and $\lambda$, these are no other adjustable free parameters. So that this
EFT will provide powerful prediction on low energy meson physics. For
example, the theoretical prediction of on-shell decay width of
$\rho^0\rightarrow e^+e^-$ is $7.0$MeV, which agree with experimental
data, $6.77\pm 0.32$MeV, very well.

In this EFT, the $\rho$-resonance propagator(Breit-Wigner formula) can be
naturally derived due to unitarity of the model instead of
input\cite{Rho},
\begin{eqnarray}\label{14}
\Delta_{\mu\nu}^{(\rho)}(q^2)=\frac{-i\delta_{\mu\nu}}{q^2-\td{m}_\rho^2
  +i\sqrt{q^2}\Gamma_{\rho}(q^2)},
\end{eqnarray}
where we have included only that part of the propagator which survives
when coupled to conserved currents, $\td{m_{\rho}}$ is the (real valued)
{\bf mass parameter} and $\Gamma_{\rho}(q^2)$ is the momentum-dependent
width,
\begin{eqnarray}\label{15}
\Gamma_\rho(q^2)=-\frac{f_\pi^2b^2(q^2)q^4}{2(1+2\zeta)^2}\sqrt{q^2}
  {\rm Im}\{\frac{\Sigma_0(q^2)}{1+\Sigma(q^2)}\}=
  \frac{|g_{\rho\pi\pi}(q^2)|^2q^4}{48\pi}\sqrt{q^2}
  (1-\frac{4m_\pi^2}{q^2})^{3/2}.
\end{eqnarray}
Numerically, the on-shell width
$\Gamma_\rho=\Gamma_\rho(q^2=m_\rho^2)=146$MeV, which agree with data
very well. 

Because the width in $\rho$-resonance (possessing a complex
pole) propagator (\ref{14}) is momentum-dependent, it must be addressed
that {\bf the mass parameter $\td{m}_\rho$ is not the physical mass
$m_\rho=770$MeV.} Let us interpret this point briefly. Empirically,
the physical mass of resonance is defined as position of pole(real value)
in relevant scattering cross section, or theoretically, it should be
defined as real part of complex pole possessed by resonance. It is
well-known that the width of $\rho$-resonance is generated by pion
loops. For a simple VMD model, the leading order of $\rho-\pi\pi$ coupling
is independent of $q^2$. Thus one has $\Gamma_\rho^{(VMD)}(q^2)\propto
\sqrt{q^2}$, and due to equation
\begin{eqnarray}\label{16}
q^2-\td{m}_\rho^2+i\frac{\Gamma_\rho}{m_\rho}q^2=0,
\end{eqnarray}
we obtain the complex pole of $\rho$-resonance is
$q^2=m_\rho^2(1-i\ep+O(\ep^2))$ with $\ep=\Gamma_\rho/m_\rho\simeq 0.19$. 
The result yields $\td{m}_\rho=m_\rho\sqrt{1+\ep^2}=784$MeV. In
particular, in the EFT used by this present paper, $\rho-\pi\pi$
coupling is proportional to $q^2$ at least. Hence one has
$\Gamma_\rho(q^2)\propto q^4\sqrt{q^2}$ at least, and complex pole
equation
\begin{eqnarray}\label{17}
q^2-\td{m}_\rho^2+i\frac{\Gamma_\rho}{m_\rho^5}q^6=0.
\end{eqnarray}
It yields $\td{m}_\rho=m_\rho\sqrt{1+3\ep^2}=810$MeV, which poses a
significant correction. 

The above discussions imply that: 1) For resonance with large width, the
mass parameter in its propagator is different from its physical mass.
The correction is proportional to the ratio of resonant width to
physical mass. 2) The mass in the orignal effective lagrangian only
emerges as a parameter instead of a physical quantity measured directly in
experiment. 3) The choice of mass parameter is relied on the choice of
model. But the physical quantity must be independent of this choice. 

Since in our result all hadronic couplings include all order information
of the chiral perturbative expansion and one-loop effects of pseudoscalar
mesons, the momentum-dependence of $\Gamma_\rho(q^2)$ is very complicate.
It is difficult to determine $\td{m}_\rho$ via the above method.
Note that it is welcome that all vector meson resonances
degenrate into a universal mass parameter $m_{_V}$ at chiral limit and
large $N_c$ limit. A reliable method is to determine $m_{_V}$ via input
mass of $\omega$-resonance (since $\Gamma_\omega\ll m_\omega$,
$\td{m}_\omega$ is almost equal to $m_\omega$ and hereafter we do not 
distingusih them). Then $\td{m}_\rho$ can be
obtained via dynamical calculation provide by this EFT. In general,   
the splitting between $\td{m}_{\rho^0}$ and $m_\omega$ is caused by three
sources: $\rho^0-\omega$ mixing, electromagnetic effects dur to VMD and
one-loop effects of pseudoscalar mesons. Up to next to leading order
$N_c^{-1}$ expansion, the momentum-dependent $\rho^0-\omega$ mixing has
been derived in ref.\cite{Omega},
\begin{equation}\label{18}
{\cal L}_{\omega\rho}^{c}=\int\frac{d^4q}{(2\pi)^4}e^{iq\cdot x}
 \Theta_{\omega\rho}(q^2)(q^2\delta_{\mu\nu}-q_\mu q_\nu)
  \omega^\mu(q)\rho^{0\nu}(x),
\end{equation}
where
\begin{eqnarray}\label{19}
\Theta_{\omega\rho}(q^2)&=&\frac{N_c}{6\pi^2}\frac{m_u-m_d}{m}
 \{g^{-2}h_0(q^2)(1-\frac{4}{3}\zeta)+q^2b(q^2)s(q^2)
   [\Sigma_K(q^2)-\frac{\Sigma_\pi(q^2)}{1+\Xi(q^2)}]\} \nonumber \\
  &&+\frac{\alpha_{\rm e.m.}\pi}{3}b_{\rho\gamma}^2(q^2)
  +O((a_1(m_u-m_d)+a_2\alpha_{\rm e.m.})^2).
\end{eqnarray}
The function $b(q^2)$, $b_{\rho\gamma}(q^2)$, $\Sigma_K(q^2)$ and
$\Sigma_\pi(q^2)$ are given in eq.(\ref{8}), and $s(q^2)$, $h_0(q^2)$
and $\Xi(q^2)$ are of follows
\begin{eqnarray}\label{20}
s(q^2)&=&\frac{4}{gf_\pi^2}
[h_0(q^2)+\frac{3}{4}g_A^2(h_1(q^2)-\frac{h_2(q^2)}{2})],\nonumber \\
h_0(q^2)&=&\int_0^1 dt\frac{6t(1-t)}{1-t(1-t)q^2/m^2}, \nonumber\\
h_1(q^2)&=&\int_0^1dt_1\cdot t_1^2\int_0^1dt_2(1-t_2) 
  \frac{3-2t_1^2t_2(1+2t_1)(1-t_2)q^2/m^2}{[1-t_1^2t_2(1-t_2)q^2/m^2]^2},
   \\
h_2(q^2)&=&\int_0^1dt_1\cdot t_1^2\int_0^1dt_2(1-t_2) 
  \frac{4(1-t_1)[3-4t_1^2t_2(1-t_2)q^2/m^2]}
   {[1-t_1^2t_2(1-t_2)q^2/m^2]^2}, \nonumber \\
\Xi(q^2)&=&4f_\pi^{-2}(1+\frac{q^2N_c}{4\pi^2f_\pi^2})\Sigma_\pi(q^2).
 \nonumber
\end{eqnarray}
In addition, it should be also noticed that the $\pi$-loop correction to
$m_\omega$ is suppressed by isospin conservation. Thus one-loop correction
to $m_\omega$ is dominated by $K$-meson. At tree level, the $\omega-KK$
coupling reads
\begin{eqnarray}\label{21}
{\cal L}_{\omega KK}&=&\frac{i}{2}
 \int\frac{d^4q}{(2\pi)^4}e^{iq\cdot x}b(q^2) 
 (q^2\delta_{\mu\nu}-q_\mu q_\nu)\omega^{\mu}(q) \nonumber \\
  &\times&\{[K^+(x)\pa^\nu K^-(x)-\pa^\nu K^+(x)K^-(x)]
  +[K^0(x)\pa^\nu\bar{K}^0(x)-\pa^\nu K^0(x)\bar{K}^0(x)]\}.
\end{eqnarray}

Then the physical mass of $\omega$-meson are
\begin{eqnarray}\label{22}
m_\omega^2&=&m_{_V}^2+{\rm Re}\{\frac{q^4\Theta_{\omega\rho}(q^2)}
 {q^2-\td{m}_\rho^2+i\sqrt{q^2}\Gamma_\rho(q^2)}
 +\frac{\pi\alpha_{\rm e.m.}}{9}q^2b_{\rho\gamma}^2(q^2)\}
 |_{q^2=m_\omega^2}      \nonumber \\
 &&-\left.
\frac{q^4b^2(q^2)\Sigma_K(q^2)}{(1+2\zeta)^2\{1+2f_\pi^{-2}
  [1+\frac{q^2C(q^2)}{1+11\zeta/3}]\Sigma_K(q^2)\}}
  \right|{\atop q^2=m_\omega^2}.
\end{eqnarray}
Input $m_\omega=782$MeV, one has $m_{_V}=785.8$MeV. Similarly, the
mass parameter of $\rho$-meson are
\begin{eqnarray}\label{23}
\td{m}_\rho^2&=&m_{_V}^2+{\rm Re}\{\frac{q^4\Theta_{\omega\rho}(q^2)}
 {q^2-m_\omega^2+im_\omega\Gamma_\omega }
 +\pi\alpha_{\rm e.m.}q^2b_{\rho\gamma}^2(q^2) 
 +\frac{f_\pi^2b^2(q^2)q^6\Sigma_0(q^2)}{2(1+2\zeta)^2(1+\Sigma(q^2))}\}
 |_{q^2=m_\rho^2}.
\end{eqnarray}
  
Using the above value of $m_{_V}$, we have $\td{m}_\rho=803.1$MeV which
is indeed significantly different from the physical mass $m_\rho=770$MeV.
Success of this prediction will be checked in the following by localizing
the position of pole in cross section of $e^+e^-\rightarrow\pi^+\pi^-$. 
Furthermore, the detail calculation shows that, the $\rho^0-\omega$ only
makes $\td{m}_\rho$ shift $-0.25$MeV, the VMD effects and one-loop effects
of pseudoscalar mesons make $\td{m}_\rho$ shift $+3.45$MeV and $+14.1$MeV
respectively. 

For working out full shape of $e^+e^-\rightarrow\pi^+\pi^-$ cross section,
the $\omega-\pi\pi$ coupling is needed. It has been derived in
ref.\cite{Omega}, in which not only the coupling via $\rho$ exchange, but
also the ``direct'' coupling are included,
\begin{eqnarray}\label{24}
{\cal L}_{\omega\pi\pi}^c=
  -i\int\frac{d^4q}{(2\pi)^4}e^{iq\cdot x}g_{\omega\pi\pi}(q^2)
  (q^2\delta_{\mu\nu}-q_\mu q_\nu)
  \omega^\mu(q)[\pi^+(x)\pa^\nu\pi^-(x)-\pa^\nu\pi^+(x)\pi^-(x)],
\end{eqnarray}
where
\begin{eqnarray}\label{25}
g_{\omega\pi\pi}(q^2)=\frac{q^2\Theta_{\omega\rho}g_{\rho\pi\pi}(q^2)}
 {q^2-\td{m}_\rho^2+i\sqrt{q^2}\Gamma_\rho(q^2)}-g_{\omega\pi\pi}^{(0)}
 (q^2),
\end{eqnarray}
with ``direct'' coupling strength
\begin{eqnarray}\label{26}
g_{\omega\pi\pi}^{(0)}(q^2)&=&\frac{N_c}{12\pi^2}\frac{m_u-m_d}{m}
  \{s(q^2)\left(\frac{1}{1+\Xi(q^2)}-\frac{10}{3}\zeta\right)
       \nonumber \\&&
   -6f_\pi^{-2}\Sigma_K(q^2)[8m^2f_\pi^{-2}b(q^2)-\frac{s(q^2)}{3}
   (1+\frac{q^2N_c}{4\pi^2f_\pi^2})]\}
   +\frac{2\alpha_{\rm e.m.}\pi}{3}\bar{F}_\pi(q^2)b_{\rho\gamma}(q^2).
\end{eqnarray}

Eqs.~(\ref{7}), (\ref{10}) and (\ref{24}) lead to the electromagnetic form
factor of pion as follow
\begin{eqnarray}\label{27}
F_\pi(q^2)=1+\frac{q^2b_\gamma(q^2)}{1+\Sigma(q^2)}-
  \frac{q^4b_{\rho\gamma}(q^2)g_{\rho\pi\pi}(q^2)}{2(q^2-\td{m}_\rho^2
  +i\sqrt{q^2}\Gamma_\rho(q^2))}-\frac{q^4b_{\rho\gamma}(q^2)
  g_{\omega\pi\pi}(q^2)}{6(q^2-m_\omega^2+i\sqrt{q^2}\Gamma_\omega)}.
\end{eqnarray}
Here due to narrow width of $\omega$, we ignore the momentum-dependence of
$\Gamma_\omega$. In this form factor, we can see that the contributions of
resonance exchange accompany $q^4$ factor. Due to this reason, some
authors declared that the pion form factor in WCCWZ EFT exhibits an
unphysical high energy behaviour($\mu>m_\rho$). However, this conclusion
is wrong. It is caused by their wrong result for momentum-dependence of
$\Gamma_\rho(q^2)$ which is fitted by experimental instead of by dynamical
prediction. In fact,
since $\sqrt{q^2}\Gamma_\rho(q^2)$ is proportional to $q^6$ at least, we 
do not need to worry that the form factor has a bad high energy behaviour. 
We can also see that there is a moment-dependent non-resonant 
contribution. It together with the contribution of resonance exchange
determined the high energy behaviour of the factor. The cross-section for
$e^+e^-\rightarrow\pi^+\pi^-$ is given by(neglecting the electron mass)
\begin{eqnarray}\label{28}
\sigma=\frac{\pi\alpha_{\rm e.m.}^2}{3}\frac{(q^2-4m_\pi^2)^{3/2}}
  {(q^2)^{5/2}}|F_\pi(q^2)|^2.
\end{eqnarray}

\begin{figure}[tph]
   \centerline{
   \psfig{figure=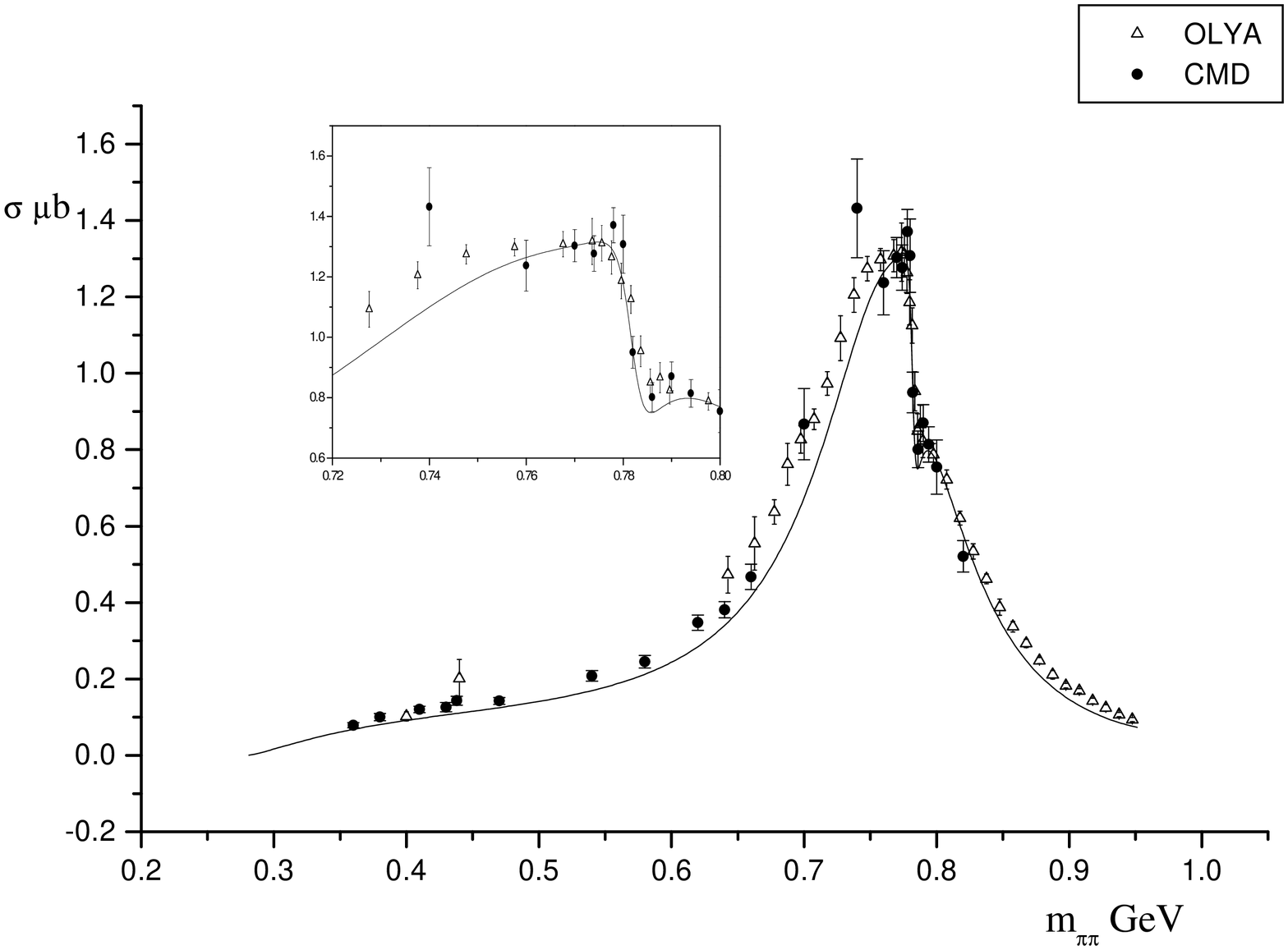,width=5.8in}}
\begin{center}
\begin{minipage}{5in}
   \caption{$e^+e^-\rightarrow\pi^+\pi^-$ cross section. The
  experimental data are from refs.[1,2].}
\end{minipage}
   \end{center}
\end{figure}

From defination of function $A(q^2)$ and $B(q^2)$ in eq.~(\ref{8}) we can
see this EFT is unitary only for $q^2<4m^2$. Thus the effective prediction
should be below $m_{\pi\pi}<2m=960$MeV. The result is shown in fig. 1. We
can see the prediction agree with data well. Especially, the theoretical
prediction in vector meson energy region agree with data excellently.
Although the mass parameter $\td{m}_\rho=803.1$MeV in $\rho$ propagator is
larger than physical mass, the position of pole is localized in
$\sqrt{q^2}=772$MeV which is just the physical mass of $\rho$. It strongly
supports our above discussion and dynamical calculation. It also
implies that we must carefully distinguish the physical mass difference of
$\rho^0$ and $\omega$ from the diffrence of mass parameter in effective
lagrangian.  

Let us give some further remarks on pion form factor~(\ref{27}). From
eqs.~(\ref{8}), (\ref{11}) and (\ref{25}) we can see that, in
eq.~(\ref{27}), $b_\gamma(q^2)$, $b_{\rho\gamma}(q^2)$, etc., are all
complex function instead of real function. It is caused by one-loop
effects of pions. Thus the expression~(\ref{27}) can be rewritten as
follow \begin{eqnarray}\label{30}
F_\pi(q^2)=1+q^2a_1(q^2)e^{i\phi_1(q^2)}-
  \frac{q^4a_2(q^2)e^{i\phi_2(q^2)}}{2(q^2-\td{m}_\rho^2
  +i\sqrt{q^2}\Gamma_\rho(q^2))}-\frac{q^4a_3(q^2)e^{i\phi_3(q^2)}}
 {6(q^2-m_\omega^2+i\sqrt{q^2}\Gamma_\omega)}.
\end{eqnarray}
Here $a_i(q^2)(i=1,2,3)$ are three real function and
$\phi_i(q^2)(i=1,2,3)$ are three momentum-dependent phases. In particular,
$\phi_3(q^2=m_\omega^2)$, so called Orsay phase, has been extracted from
data as $100-125$ degrees\cite{Bena97,Connell96}. Our theorectical
prediction is $\phi_3(q^2=m_\omega^2)=116.5$ degrees. However, so far, the
phases $\phi_1(q^2)$ and $\phi_2(q^2)$ are not reported in any
literatures. These momentum-dependent phases indicate that the dynamics
including loop effects of pseudoscalar mesons is different from one only
in tree level. In fig.2, we given theorectical curves of $\phi_i(q^2)$.
They are indeed nontrivial.

\begin{figure}[hptb]
  \centerline{\psfig{figure=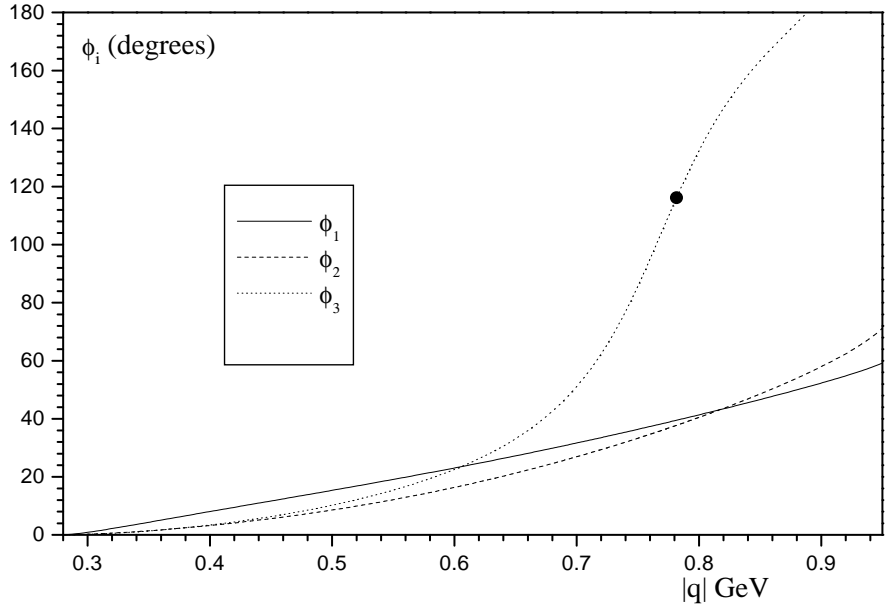,width=5.5in}}
\begin{center}
\begin{minipage}{5in}
   \caption{$\phi_i$ versus $m_{\pi\pi}$ in GeV. Here the solide line
denotes the phase shift of non-resonant background $\phi_1$, the dash line
denotes the phase shift $\phi_2$ in $\rho$ coupling and the dot line
denotes the phase shift $\phi_3$ in $\omega$ coupling. ``$\bullet$''
denotes Orsay phase.}
\end{minipage}
   \end{center}
\end{figure}

Obviously, $F_\pi(q^2)$ is an analytic function in the complex $q^2$
plane, with a branch cut along the real axis beginning at the two-pion
threshold, $q^2=4m_\pi^2$. Time-reversal invariance and the unitarity of
the $S$-matrix requires that the phase of the form factor be that of
$l=1,\;I=1$ $\pi-\pi$ scattering\cite{GT58}. This last emerges as
$\pi-\pi$ scattering in the relevant channel is very nearly elastic from
threshold through $q^2\simeq (m_\pi+m_\omega)^2$\cite{HL81,HOP73}. In this
region of $q^2$, then, the form factor is related to the $l=1,\;I=1$
$\pi-\pi$ phase shift, $\delta_1^1$, via\cite{Gasi66}
\begin{eqnarray}\label{31}
F_\pi(q^2)=e^{2i\delta_1^1(q^2)}F_\pi^*(q^2),
\end{eqnarray}
so that
\begin{eqnarray}\label{32}
\tan{\delta_1^1(q^2)}=\frac{{\rm Im}F_\pi(q^2)}{{\rm Re}F_\pi(q^2)}.
\end{eqnarray}
The above is a special case of what is sometimes called the
Fermi-Watson-Aidzu phase theorem\cite{Gasi66,Watson55}. In fig. 3 and
fig. 4 we plot theoretical curves of the $l=1,\;I=1$ $\pi-\pi$ phase shift 
$\delta_1^1$ versus $m_{\pi\pi}$ and of $\sin{\delta_1^1}/p_\pi^3$ versus
$m_{\pi\pi}$ (where $p_\pi=\frac{1}{2}\sqrt{q^2-4m_\pi^2}$) respectively.
We omit the $\omega$ contribution from our plots of the phase of
$F_\pi(q^2)$ for comparing with time-like region pion form factor 
data~\cite{Data73,Data74,Data77}. We have also assumed that $\delta_1^1$
is purely elastic in the regime shown, i.e., the loop effects of
$\omega-\pi$ are omitted. The curve predicts
$\delta_1^1\rightarrow 90^\circ$ as $\sqrt{q^2}\rightarrow 774{\rm MeV}
\simeq m_\rho$, and $\delta_1^1>100^\circ$ for $\sqrt{q^2}>787$MeV. These
results agree with data very well.

\begin{figure}[hptb]
  \centerline{\psfig{figure=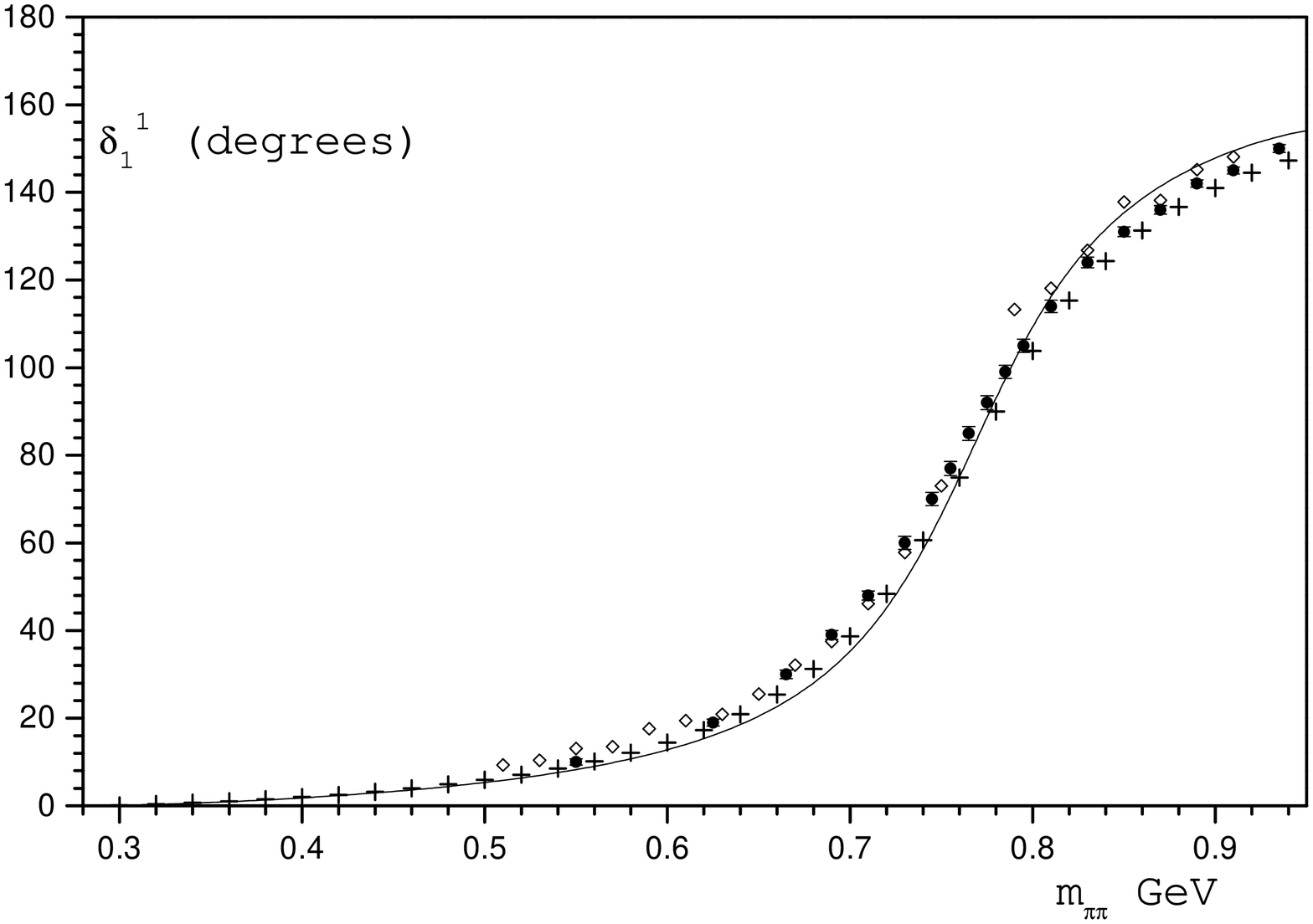,width=5.5in}}
\begin{center}
\begin{minipage}{5in}
\caption{The $l=1,\;I=1$ $\pi-\pi$ scattering phase shift. The solid
circle point are these from [22], the hollow diamond point are from
[23] and ``+'' denotes the points from [24]. Note that the $\omega$
contribution to the time-like pion form
factor phase has been omitted, to facilitate comparison with the
empirical phase shift.} 
\end{minipage}
   \end{center}
\end{figure}
  
Finally we discuss the near threshold behaviour of the form factor.
1) The chiral perturbative theory predicts the form factor at
threshold to be $[F_\pi(4m_\pi^2)]_{\rm ChPT}=1.17\pm 0.01$, and ours,
$[F_\pi(4m_\pi^2)]=1.154$, is close to the ChPT result.
2) The electromagnetic radius of charged pion has been determined to
be $\sqrt{<r^2>_\pi}=0.657\pm 0.027$fm\cite{Dally82}, whereas the
theorectical prediction in this present paper is $\sqrt{<r^2>_\pi}=0.635$
fm. 3) The Froggatt-Petersen phase shift function
$\sin{\delta_1^1}/p_\pi^3$ is connected with the vector-isovector
$\pi-\pi$ scattering length $a_1^1$ through
\begin{eqnarray}\label{33}
a_1^1=\lim_{q^2\rightarrow 4m_\pi^2}\frac{\sin{\delta_1^1}}{p_\pi^3}.
\end{eqnarray}
Our theoretical prediction is $a_1^1=0.037$ in unit of $m_\pi^{-3}$. This
value is very close to experimental results from $K_{e4}$
data\cite{Nagel79,Rosselet77} using a Roy equation fit ($a_1^1=0.038\pm
0.002$) and ChPT prediction $a_1^1=0.037\pm 0.01$\cite{Knecht95} at the
two loop order (at $O(p^4)$).

In summary, a rigorous, unitary EFT method has been applied to study pion
form factor and $l=1,\;I=1$ $\pi-\pi$ phase shift. The theoretical
predictions include all order informations of the chiral perturbative
expansion and one-loop effects of pseudoscalar mesons. The Breit-Wigner
formula for resonant propagators is derived by the EFT itself instead of
an input. The momentum-dependence of $\Gamma_\rho(q^2)$ is predicted by
the dynamics. It also has been revealed that the mass parameter in
resonant propagators should be different from its physical mass due to
momentum-dependent width. This point is confirmed by both of dynamical
calculation and phenomenological fit. It also tells us how to
understand the mass splitting between $\rho^0$ and $\omega$: Although the
dynamical calculations show that the mass parameter of $\rho$ in effective
lagrangian is even larger than one of $\omega$, the position of pole
localized in real aixs give their right physical mass splitting. 
The contribution to this mass splitting from $\rho^0-\omega$ mixing is
very small, the dominant contribution is from one-loops effects of
pseudoscalar mesons. The EFT mechanics on $\rho^0-\omega$ mass splitting
revealed in the present paper is rather subtle. And it is another evident
to confirm again that the EFT of QCD proposed in ref.\cite{Rho} is sound.
Actually, to the best of our knowledges, this is the first time to get
$\rho^0-\omega$ mass splitting through a well-defined quantum field theory
calculation.

Due to one-loop effects of pions, the photon-$\pi\pi$, photon-vector and
vector-$\pi\pi$ coupling are all with a nontrivial phase shift instead of
purely real in some simple models. In a series of recent papers, we have
revealed that the one-loop effects of pseudoscalar mesons play a very
important role in low energy hadronic physics. Theoretically, it keeps
unitarity of the $S$-matrix, and numerically, its contributions are about
$30\%$. A well-defined EFT must be able to evaluate the high order
contributions of the chiral perturbative expansion and $N_c^{-1}$
expansion. It is an important criterion to judge a model as a rigorou EFT
or a phenomenological model only.

In our study, no parameters need to be fitted by the data of pion form
factor and $l=1,\;I=1$ $\pi-\pi$ phase shift. Thus our results are
rigorous theoretical predictions and agree with data very well. 

\begin{figure}[hptb]
  \centerline{\psfig{figure=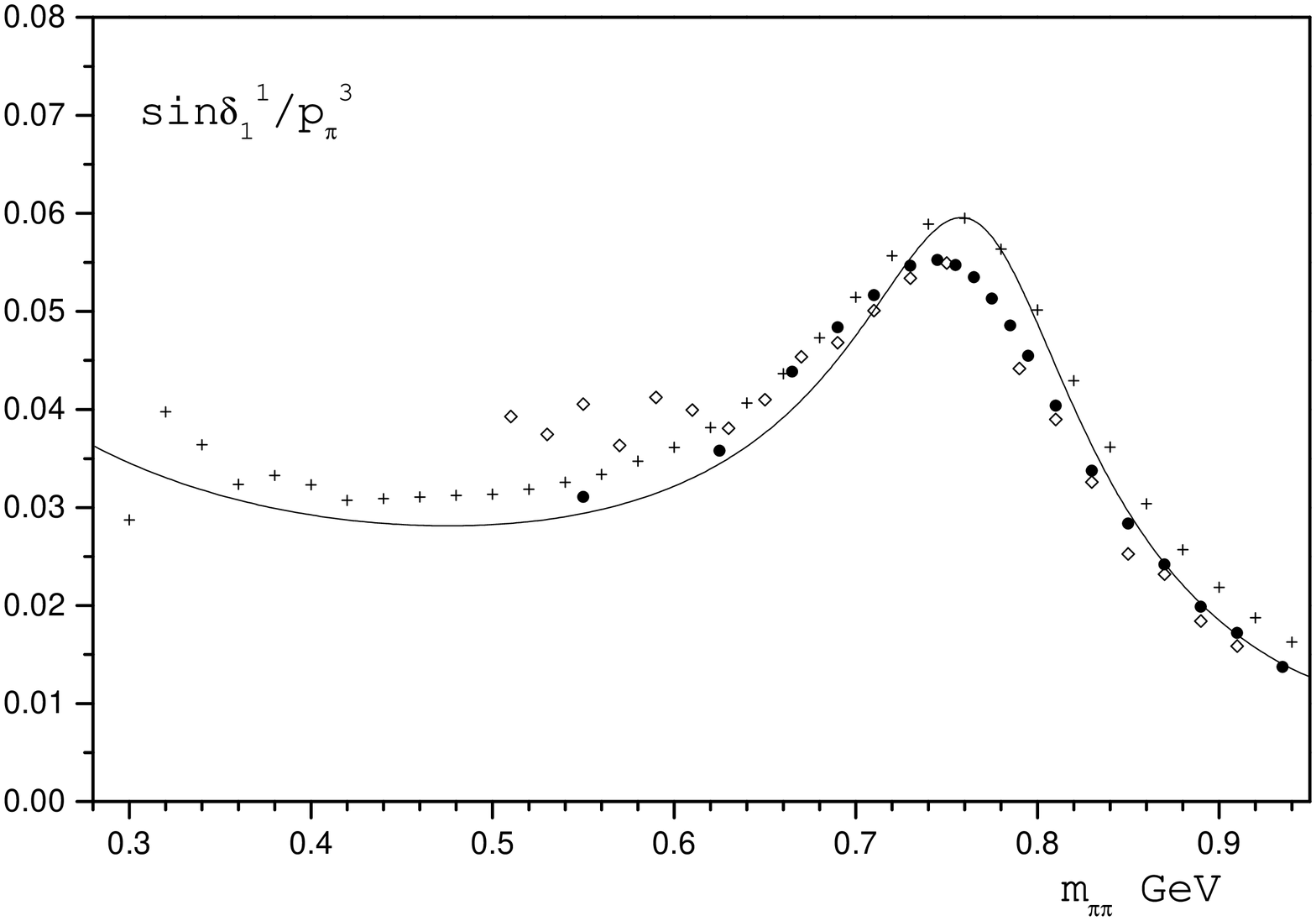,width=5.5in}}
\begin{center}
\begin{minipage}{5in}
\caption{Function $\sin{\delta_1^1}/p_\pi^3$ deduced from $\pi^+\pi^-$
phase shift; the function is given in units of $m_\pi^{-3}$. The solid
circle point are these from [22], the hollow diamond point are from
[23] and ``+'' denotes the points from [24].}
\end{minipage}
   \end{center}
\end{figure}

\end{document}